\RequirePackage{fix-cm}
\documentclass[smallextended]{svjour3}       % onecolumn (second format)
\smartqed  % flush right qed marks, e.g. at end of proof
\usepackage{graphicx}
\usepackage{amsmath,amssymb}
\usepackage{eurosym}

\begin{document}

\title{Quantum Image Rotation by an arbitrary angle%\thanks{Grants or other notes
}


\author{Fei Yan \and Kehan Chen \and Salvador E. Venegas-Andraca \and Jianping Zhao}


\institute{Fei Yan. \at School of Computer Science and Technology, Changchun University of Science and Technology.
              No. 7089, Weixing Road, Changchun 130022, China.
           \and Kehan Chen. \at School of Computer Science and Technology, Changchun University of Science and Technology. No. 7089, Weixing Road, Changchun 130022, China.
           \and Salvador E. Venegas-Andraca. \at Tecnol\'{o}gico de Monterrey, Escuela de Ingenier\'{i}a y Ciencias and Campus Estado de M\'exico. Carretera al Lago de Guadalupe KM. 3.5, Atizap\'{a}n de Zaragoza, Estado de M\'{e}xico CP 52926, M\'{e}xico.
           \email{salvador.venegas-andraca@keble.oxon.org}
           \and Jianping Zhao. \at School of Computer Science and Technology, Changchun University of Science and Technology, No. 7089, Weixing Road, Changchun 130022, China.
}

\date{Received: date / Accepted: date}

\maketitle

\begin{abstract}
In this paper, a novel method of quantum image rotation (QIR) based on shear transformations on NEQR quantum images is proposed. To compute the horizontal and vertical shear mappings required for  rotation, we have designed quantum self-adder, quantum control multiplier, and quantum interpolation circuits as the basic computing units in the QIR implementation. Furthermore, we provide several examples of our results by presenting computer simulation experiments of QIR under $30^\circ$, $45^\circ$, and $60^\circ$ rotation scenarios and have a discussion onto the anti-aliasing and computational complexity of the proposed QIR method. 
\end{abstract}

\section{Introduction}
\label{intro}

Cross-pollination between physics and computer science has long been abundant and fruitful. For example, statistical mechanics has inspired the development of probabilistic algorithms focused on solving optimization problems \cite{kirkpatrick83} while a novel approach in physics is to think of natural phenomena in terms of computational procedures (e.g., \cite{cubitt15}).

Since the publication of the seminal works by Feynman \cite{Feynman1986qmcomputers,Feynman1982Simulating}, the sustained efforts of scientists and engineers have provided the grounds for the fields of quantum computation and quantum information (QC \& QI) to transition from emerging branches of science into mature research fields of science and engineering. As a result and in addition to further consolidation of the mathematical and physical foundations of QC \& QI, recent initiatives include the development of quantum algorithms and quantum communication protocols in fields like artificial intelligence \cite{quail}, machine learning \cite{quantum-machine-learning}, computational geometry \cite{computational-geometry}, military technology \cite{quantum-radar}, and image processing \cite{seva_special_issue_QIMP2015,special_issue_QIMP2015}. Furthermore, the development of incipient branches of industry and high-tech business based on quantum technology \cite{idquantique,dwave,ibm,microsoft,googlequantum,1qbit,rigetti,qmanifesto,ukquantumhubs,qis_executive_usa,theeconomistquantum2017,mittechreviewquantum2017}, roadmaps of quantum technology \cite{qist04,era07,ukqt15} as well as an increasing presence in the arena of intellectual property \cite{winiarczyk13,ukquantumtech14,ribordy09,berkley11,troyer15,hunt16} make QC \& QI strong candidates to create new advanced technology markets and, consequently, key features for further economic development in the $21^\text{st}$ century.
  
Quantum image processing (QIMP), an emerging sub-discipline  of QC \& QI, is a field devoted to capturing, manipulating, and recovering visual information using quantum mechanical systems.  QIMP was born with the publication of \cite{vlasov,beach,sva01,sva02} and, since then, it has amassed a spurt of interest from researchers with diverse backgrounds who have advanced this field by proposing quantum algorithms for image storage and retrieval \cite{Yan2014Quantum,Yan2016A}, image encryption/decryption \cite{image-encryption-decryption}, image segmentation \cite{image-segmentation,Yan2012Quantum}, image watermarking \cite{Yan2015A} and image filtering \cite{image-filtering}, as well as quantum operations to manipulate quantum images like geometric transformations \cite{Le2010Fast}, image comparison \cite{Yan2013A} and image translation \cite{Wang2015Quantum}, among many other contributions.  The advancement of QIMP brought the development of more intuitive and flexible representations of quantum images, among them the Flexible Representation of Quantum Images (FRQI)  \cite{Le2011A} as well as a FRQI-based, novel enhanced quantum representation (NEQR) \cite{Zhang2013NEQR} in which grayscale pixel values are stored on a sequence of qubits. For full reviews of quantum image representation models and emergent branches of quantum image processing, the reader is referred to \cite{Yan2016A,aburaed17}.

In classical image processing and related fields like computer vision and pattern recognition, image rotation is regarded as a key tool for image registration, image fusion, and image mosaicing, for instance. Nonetheless, image rotation has not  been sufficiently studied within the quantum computing framework. To fill this gap, we propose in this paper a quantum algorithm for image rotation (QIR) consisting of a sequence of three shear mappings (horizontal, vertical, then horizontal again) onto NEQR images. 

Our contribution is composed of the following stages:

\begin{description}
  \item[(\romannumeral1)] In classical image processing, it is known that an arbitrary 2D rotation can be performed by a series of three (horizontal, vertical, horizontal) shear transformations  \cite{Paeth1990A,Unser1995Convolution} , i.e.
  
 \begin{equation}\label{rot_as_shear}
\begin{pmatrix}
\cos(\theta)&&\sin(\theta)\\-\sin(\theta)&&\cos(\theta)
\end{pmatrix} =
\begin{pmatrix}
1&&\alpha\\0&&1
\end{pmatrix} 
\begin{pmatrix}
1&&0\\\beta&&1
\end{pmatrix} 
\begin{pmatrix}
1&&\alpha\\0&&1
\end{pmatrix} 
\end{equation}
  
where $\begin{pmatrix} 1&&\alpha\\0&&1 \end{pmatrix} $ is a horizontal shearing transformation and $\begin{pmatrix} 1&&0\\\beta&&1 \end{pmatrix}$ is a vertical shearing transformation. A shear transformation can be defined as a transformation in which all points along a given line $\mathcal{L}$ remain fixed, while other points are shifted parallel to $\mathcal{L}$ by a distance proportional to their perpendicular distance from $\mathcal{L}$ \cite{sharma11,ShearTransfWolfram}. Moreover, the shear factor is defined as the (constant) distance a point $P$ moves due to a shear divided by the perpendicular distance of P from $\mathcal{L}$ \cite{sharma11,ShearFactorWolfram}.
 
In this paper, quantum images are partitioned into two halves by a reference line, hence obtaining top-bottom or left-right sub-images, and we compute shear transformations on both halves.  As a result, the computation of each shear transformation displaces quantum pixels in different halves to opposite directions (which is also the expected behavior of shear transformations on classical pixels).
 
\item[(\romannumeral2)] Shear transformations are linear mappings displacing each pixel towards a fixed direction by a proportional amount to its signed distance according to the shear factor. Indeed, as Eq. (\ref{rot_as_shear}) shows, shear transformations are not unitary transformations. Nevertheless, it is known that any classical irreversible circuit can be replaced by a reversible circuit that uses Toffoli gates; moreover, a Toffoli gate can be implemented both as classical and quantum logic gate, hence any classical circuit (either reversible or irreversible) can be substituted by a quantum circuit that performs the same computation \cite{Nielsen2000Quantum}. In this paper, we present a quantum circuit that computes shear transformations on quantum images, i.e. we present a unitary version of shear transformations. To do this, we suggest some modifications onto the quantum multiplier proposed in \cite{Vedral1996Quantum}.
  
\item[(\romannumeral3)] The shear factor is always defined by a trigonometric function of an angle and, consequently, pixel displacement produced by shear operations is usually expressed as floating point numbers. This is in accordance with Eq.(\ref{rot_as_shear}) as, in general, rotations would produce pixel positions described by vectors $\boldsymbol{y}_i \in \mathbb{R}^2$. Now, a well known problem in classical image processing \cite{tanimoto12} arises: since pixel positions in a digital image are described by vectors $\boldsymbol{x}_i \in \mathbb{Z}^2$, we need to define a mathematical procedure to 
accommodate $\boldsymbol{y}_i$ in $\boldsymbol{x}_i$. This problem is also faced in quantum image processing as in all known quantum image representations \cite {Yan2016A}, the {\it position} of quantum pixels is also described by elements of $\mathbb{Z}^2$. In this paper, we propose a quantum interpolation module to solve this problem.
\end{description}

The following content is organized as follows. The NEQR representation and the quantum computing tools for shear mappings are presented in Section \ref{sec2}. The QIR operations based on the shear mappings are proposed in Section \ref{sec3}, where the quantum circuits to realize these operations are designed and discussed. In Section \ref{sec4}, the proposed work is verified by the simulation experiments as an flexible and effective tool in quantum computing domain. The conclusions of the research and the further work are located in Section \ref{sec5}.

\section{NEQR representation and quantum computing tools for shear mapping}
\label{sec2}

A quantum computer system can be viewed as a quantum network consisting of quantum logic gates, wherein each gate performs an elementary unitary operation on one, two, or more two-state quantum systems \cite{sva02}. In this section, we introduce the NEQR model and present the quantum circuits that will be used to compute shear transformations on NEQR images.

\subsection{Novel enhanced quantum image representation}
\label{sec2-1}

The Novel Enhanced Quantum Representation (NEQR) \cite{Zhang2013NEQR} is an image representation model that uses the basis states of a qubit sequence to store the grayscale value of every pixel. NEQR images are mathematically defined as follows:

\begin{equation}\label{eq1}
\begin{aligned}
|I\rangle &= {1\over 2^n}\sum_{y=0}^{2^n-1}\sum_{x=0}^{2^n-1}|f(y,x)\rangle|yx\rangle,
\end{aligned}
\end{equation}
where $|f(y,x)\rangle$ encodes the grayscale value (i.e., a number in $[0,255]$) of each pixel: %information of the image whose range of $[0,255]$ is formalized as:
\begin{equation}\label{eq2}
|f(y,x)\rangle=\mathop{\otimes}\limits_{i=0} ^{7}|C_{yx}^i\rangle=|C_{yx}^{7}C_{yx}^{6}\cdots C_{yx}^{0}\rangle, C_{yx}^k \in \{0,1\}.
\end{equation}

\begin{figure}
\centerline{\includegraphics[width=14cm]{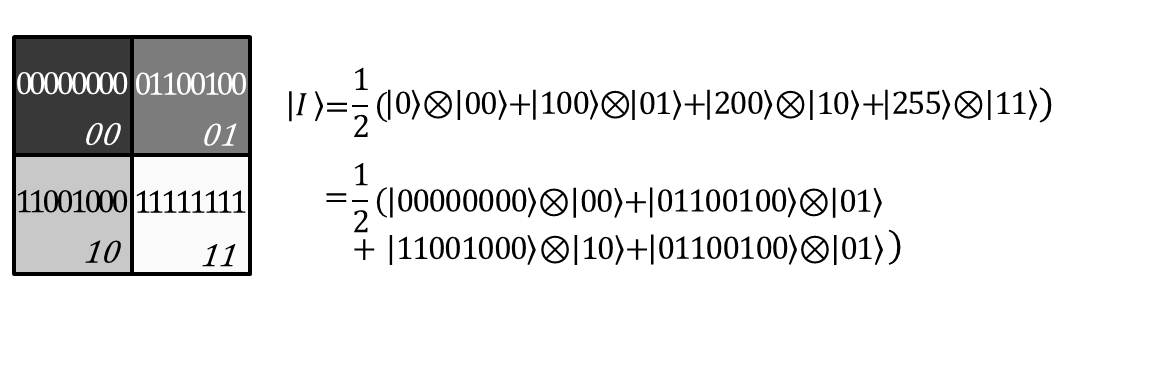}} %100 percent
\vspace*{13pt}
\caption{A $2\times 2$ NEQR image and its quantum state}
\label{fig1}
\end{figure}

An example of a $2\times 2$ NEQR image and its quantum state is presented in Fig. \ref{fig1}.

%\addfigure{fig1}{A $2\times 2$ NEQR image and its quantum state}{fig1.pdf}{width=0.80\textwidth}

\subsection{\textbf{Quantum tools to realize addition and self-addition}}
\label{sec2-2}

%\begin{figure}
%  \includegraphics{example.eps}
%\end{figure}

Either classical or quantum, designing sophisticated algorithms is a difficult task. A widely used approach in computer science, computer engineering and software development is to produce (hardware of software) modules for performing repetitive/basic tasks, so that scientists and engineers can focus on abstract problem solving using computer systems and software as tools, rather than having to deal with both the problem at hand and the practical implementation of basic building blocks for computer programming. 

Following the rationale described above, we present two basic modules, quantum adder and quantum self-adder.

\begin{description}

\item[(\romannumeral1)] Quantum adder

In this subsection, we present a quantum adder circuit as originally  introduced in \cite{Vedral1996Quantum}. The aim is to perform the following computation:

\begin{equation}\label{eq3}
|a,b\rangle \rightarrow |a,a+b\rangle
\end{equation}
where $|a\rangle $ and $|b\rangle$ are two input quantum kets and the two output kets are $|a\rangle $ and $|d\rangle $, in which $|d\rangle=|a\rangle+|b\rangle$. As presented in Fig. \ref{fig2}, a quantum adder consists of $2n-1$ carry modules and $2n$ sum modules. In addition, the carry module could be decomposed to 2 Toffoli gates and 1 C-NOT gates, while the sum module could be executed by 2 C-NOT gates as presented in Fig. \ref{fig2}(a) and Fig. \ref{fig2}(b). Moreover, as discussed in \cite{Draper2000Addition} and \cite{Vedral1996Quantum}, quantum subtraction could be implemented by the quantum adder(s) due to the fact that quantum gates are reversible. We illustrate the subtraction by locating the black bar at the left side of the module from the original right within the adder.

%\addfigure{fig2}{\textbf{The quantum adder circuit}}{fig2}{width=1.0\textwidth}
\begin{figure} [htbp]
%\vspace*{13pt}
\centerline{\includegraphics[width=12cm]{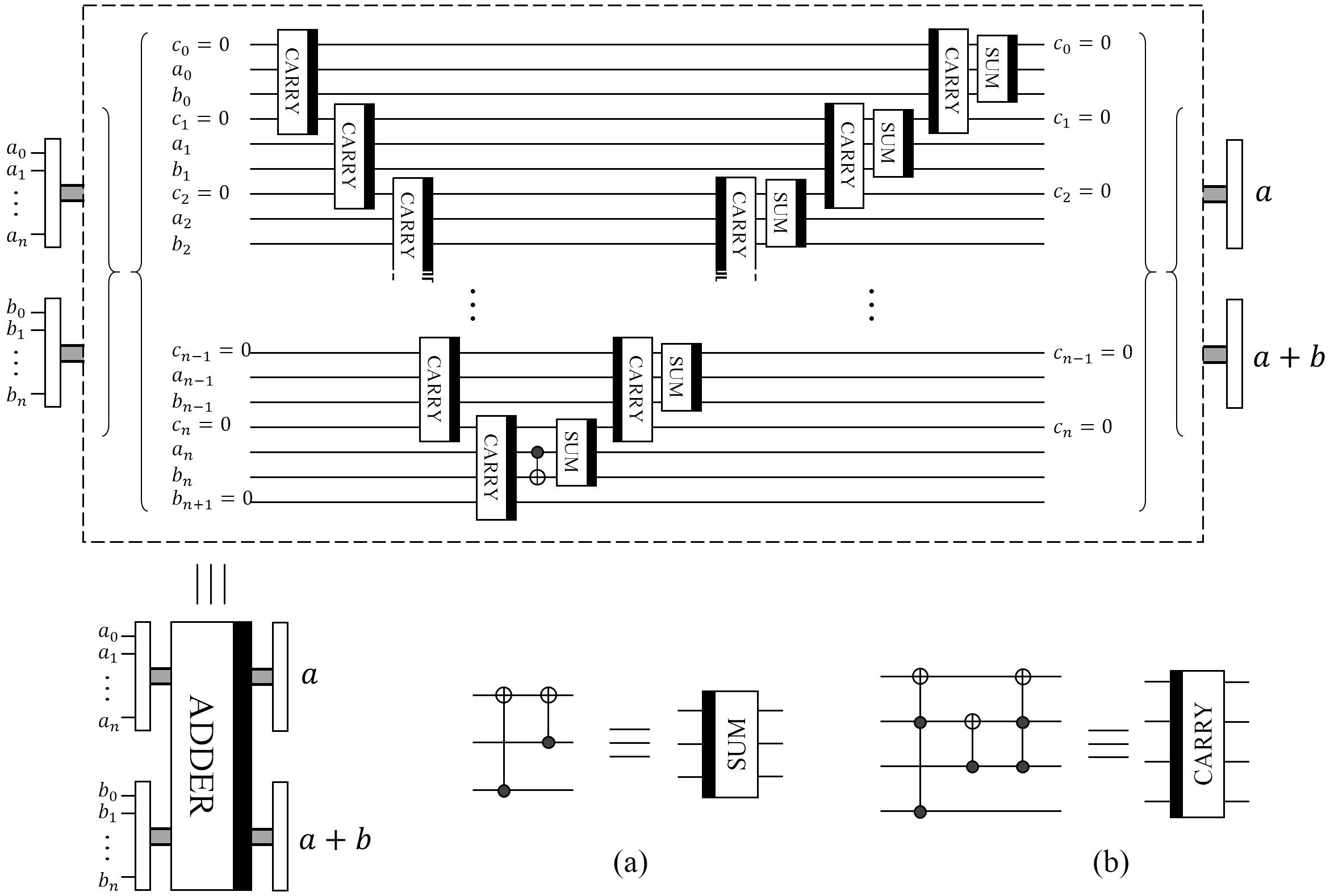}} %100 percent
\vspace*{13pt}
\caption{The quantum adder circuit}
\label{fig2}
\end{figure}

\item[(\romannumeral2)] Quantum self-adder

Let $x$ be a binary number. Then, the binary representation of $2x$ can be achieved simply by concatenating $x$ with a zero at the least significant position (which is equivalent to shifting $x$ to the left and setting the least significant position to zero). Formally, 

$$ x = \sum_{i=0}^{2^n-1} \alpha_i 2^i \Rightarrow 2x = \sum_{j=0}^{2^{n}} \beta_j 2^j$$ where $\beta_0 = 0$ and $\beta_j = \alpha_{i-1}$ for $j \in \{1, \ldots, n \}$.  Inspired by this procedure, we propose the following computation:

\begin{equation}\label{eq5}
U_{n-1} \otimes U_{n-2}\dots \otimes U_{0}\otimes I |x\rangle \otimes (\vert \overbrace{00\dots 0}^{n+1}\rangle)
= |x\rangle \otimes \vert x_{n-1}x_{n-2}\dots x_0{0}\rangle = \vert x\rangle \otimes \vert 2x\rangle
\end{equation}

where unitary operators $U_{n-1}U_{n-2}\dots U_{0}$ are  C-NOT quantum gates.
For instance, let $\vert x\rangle=\vert x_2x_1x_0\rangle=\vert 110\rangle$, then $\vert 2x\rangle=\vert 1100\rangle$, through which $U_2$=NOT, $U_1$=NOT, and $U_0$=\emph{I}.

The quantum circuit named ``SELF-ADDER" and labelled as ``S-A" presented in Fig.~\ref{fig3} implements Eq. (\ref{eq5}).

%\addfigure{fig3}{The quantum self-adder circuit}{fig3}{width=0.9\textwidth}
\begin{figure} [htbp]
%\vspace*{13pt}
\centerline{\includegraphics[width=11cm]{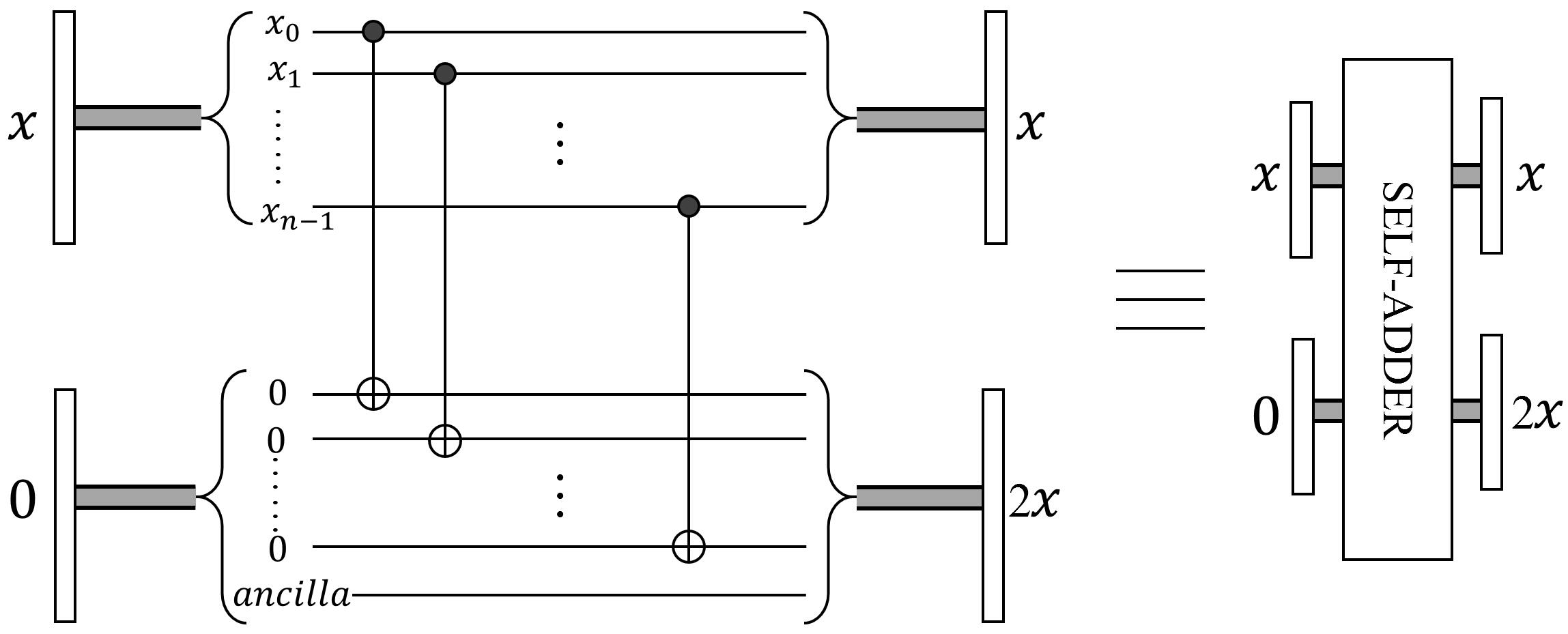}} %100 percent
\vspace*{13pt}
\caption{The quantum self-adder circuit}
\label{fig3}
\end{figure}

\end{description}

\subsection{Quantum controlled multiplier}
\label{sec2-3}
%of course
Let $a,x$ be binary numbers where $x = \sum_{i=0}^{2^n-1} 2^i x_i $. Then, the multiplication $ax$ can  be expressed as

\begin{equation} 
ax =  a  \sum_{i=0}^{2^n-1} 2^i x_i  =  \sum_{i=0}^{2^n-1} (2^i a)( x_i)
\label{sum_mult}
\end{equation}

For instance, let $a=10101$ and $x = 1011$, where $x_3 = 1, x_2 = 0, x_1 = 1, x_0 = 1$. Then $ ax = (10101) \times 
(1011)= 11100111$. Now, using  Eq. (\ref{sum_mult}):

\begin{eqnarray}
ax &=& \sum_{i=0}^3 (2^i a)( x_i) = (2^3 a)(x_3) +  (2^2 a)(x_2) + (2^1 a)(x_1) + (2^0 a)(x_0) \nonumber \\
&=& \overbrace{(1000)}^{2^3}\overbrace{(10101)}^a \overbrace{(1)}^{x_3} + \overbrace{(100)}^{2^2}\overbrace{(10101)}^a \overbrace{(0)}^{x_2} + \overbrace{(10)}^{2^1}\overbrace{(10101)}^a \overbrace{(1)}^{x_1} + \overbrace{(1)}^{2^0}\overbrace{(10101)}^a \overbrace{(1)}^{x_0} \nonumber\\
&=& 10101000 + 101010 + 10101 = 11100111.
\end{eqnarray}

We propose a quantum controlled multiplier (Ctrl-MULTI) that, based on the self-adder circuit presented above and Eq. (\ref{sum_mult}), computes the following operation:

\begin{equation}
\text{Ctrl-MULTI} |a\rangle |b\rangle |0\rangle =  |a\rangle |b\rangle |ab\rangle
\end{equation}

\begin{figure} [htbp]
%\vspace*{13pt}
\centerline{\includegraphics[width=13cm]{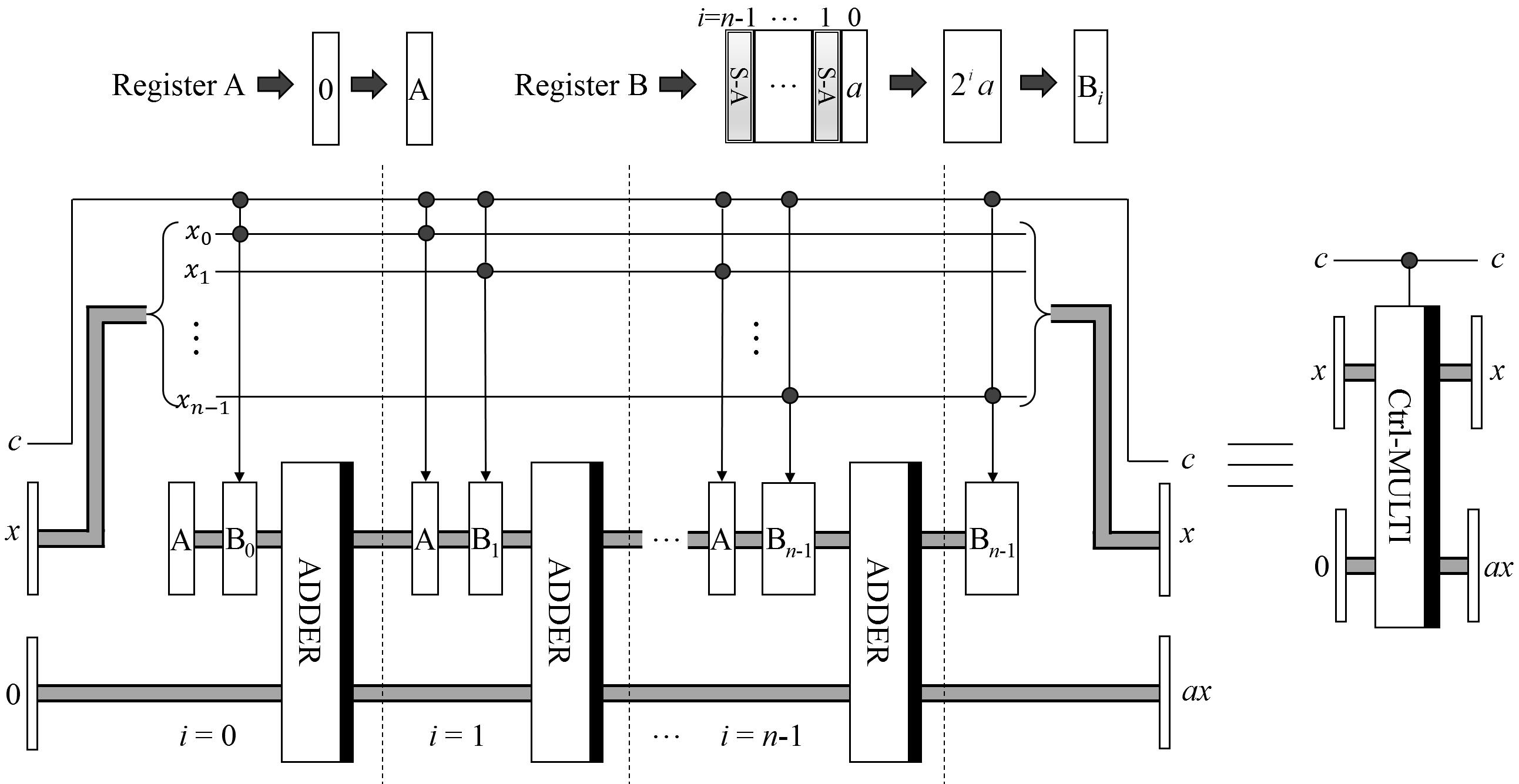}} %100 percent
\vspace*{13pt}
\caption{The circuit to realize quantum controlled multiplier}
\label{fig4}
\end{figure}

Our proposed circuit for implementing Ctrl-MULTI is presented on Fig. \ref{fig4}. Our proposal for Ctrl-MULTI is realized by \textit{n} stages of quantum adders. During each stage, it needs to be considered that whether $2^{i}a$ should be added according to the state of the qubit $|x_i\rangle, i = 0,1,\dots n-1$.

Inputs that are encoded in binary form for the computational basis of the selected qubits are called a quantum register, or simply a register. For instance, if the number 5 is loaded into a quantum register, we need to prepare three qubits in the state of $|1\rangle \otimes |0\rangle \otimes |1\rangle$. Given one quantum control multiplier, two registers (denoted as Register A and Register B) are required during the three treating steps as follows (Fig. \ref{fig4}):

Step 1: Initialize Registers A and  B as $\vert 0\rangle^{\otimes n}$ and $\vert a\rangle$, respectively. A Toffoli gate controlled by ancilla qubit ``\emph{c}" and $x_0$ is responsible for manipulating Register A or B and is taken as an input of the quantum adder in the step when $i=0$. The other input is set as $\vert 0\rangle$ temporarily.
\\

Step 2: Updating Register B to $2a$ by executing the ``Self-Adder" module in the step when $i=1$. Similarly, the Toffoli gate that is controlled by the ancilla qubit ``\emph{c}" and $x_1$ (in this turn) takes Register A or B as an input of the quantum adder. During the addition, the other input is the temporary addition outcome through the previous step, i.e. $i=0$.
\\

Step 3: Following the rationale described in steps 1 and 2, go through the circuit stages defined for $i \in \{2, \ldots , n-1\}$. The output of the last quantum adder is $ax$.

%Iterating the operations till the step when $i=n-1$. 

%\addfigure{fig4}{The circuit to realize quantum control multiplier}{fig4}{width=1.1\textwidth}

\subsection{Quantum Interpolation}
\label{sec2-4}

\begin{figure} [htbp]
%\vspace*{13pt}
\centerline{\includegraphics[width=12cm]{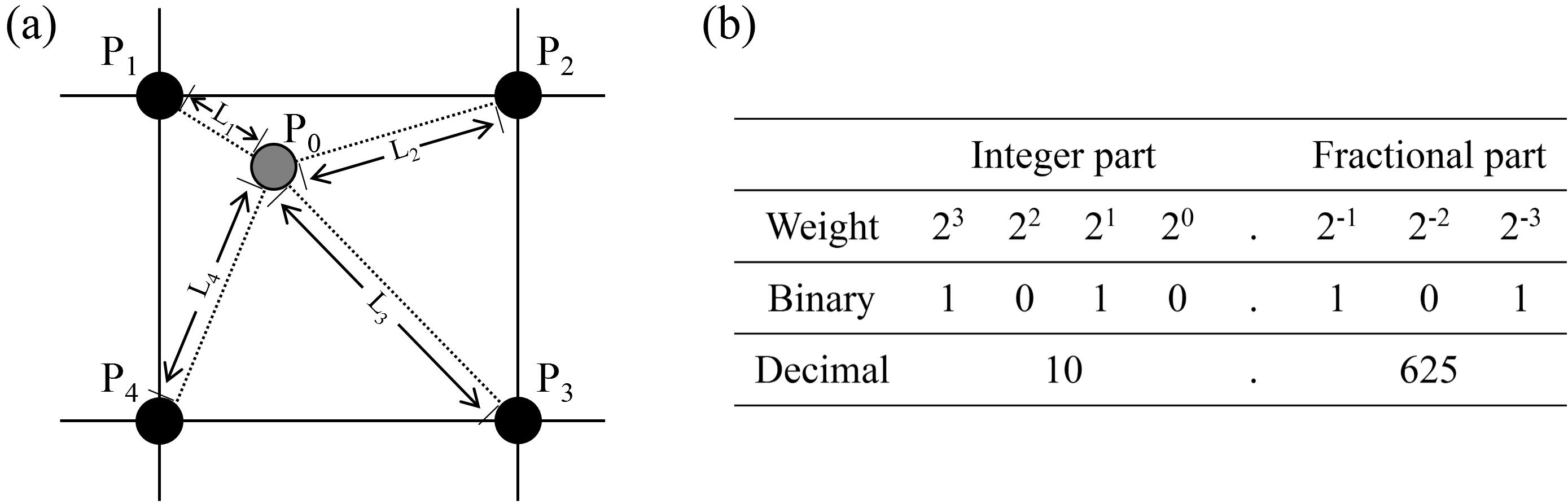}} %100 percent
\vspace*{13pt}
\caption{The principle of NNV and the conversion of decimal to binary notation}
\label{fig5}
\end{figure}

As stated earlier, the shear factor, which is essentially a trigonometric function of rotation angle, is seldom an integer. In order to locate the sheared pixel in the coordinate system, the nearest neighbor value (NNV) interpolation \cite{Rukundo2012Nearest} according to the smallest absolute difference to the four known adjacent position values is employed to determine a proper pixel position. As shown in Fig. \ref{fig5}(a), suppose that $P_1$, $P_2$, $P_3$, and $P_4$ are four adjacent pixels within one image, $P_0$ is the calculated position of a sheared pixel point through the mapping operation. To locate $P_0$ in the coordinate system, the nearest point among $P_1$, $P_2$, $P_3$, and $P_4$ should be found with the algorithm. Since $L_1<L_2<L_4<L_3$ as indicated in Fig. \ref{fig5}(a), $P_1$ will be the final position of $P_0$ after shear mapping.

%\addfigure{fig5}{The principle of NNV and the conversion of decimal to binary notation}{fig5}{width=1.0\textwidth}

In addition, in the classical binary arithmetic, the conversion from decimal to binary notation concerns the binary positions and weights. A simple example is shown in Fig. \ref{fig5}(b), where $(10.625)_{10}$ is transformed to $(1010.101)_{2}$. To cope with the infinite decimals, we keep 4 decimal places for rounding off the fraction in further to produce an integer. In binary system, if the first bit after the decimal point is ``0", it indicates the fractional part of this number is less than 0.5 and the omitting procedure is needed. Otherwise, the interpolation operation will add ``1" to its integer part. The decimal illustrated in Fig. \ref{fig5}(b) should become 11 after the interpolation operation.
%be treated into 11

\begin{figure} [htbp]
%\vspace*{13pt}
\centerline{\includegraphics[width=11cm]{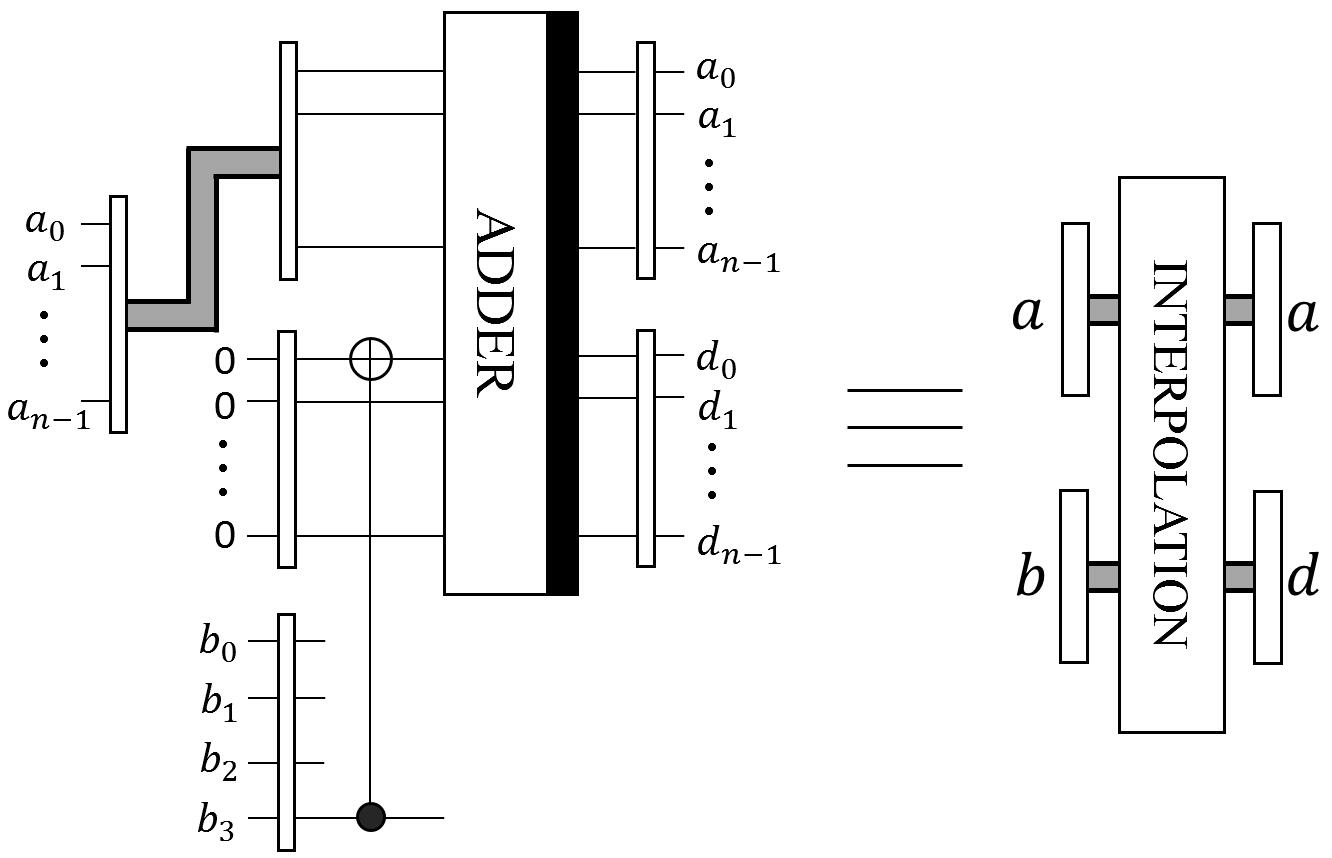}} %100 percent
\vspace*{13pt}
\caption{The circuit to realize quantum interpolation}
\label{fig6}
\end{figure}

Given the NNV interpolation in quantum computing framework, the circuit design is shown in Fig. \ref{fig6} in which $|a\rangle=|a_{n-1}\dots a_{0}\rangle$ is the integer part of a decimal and $|b\rangle=|b_3b_2b_1b_0\rangle$ is the fractional part with 4 effective decimal places. $|b_3\rangle$ is the most critical bit during converting a decimal to an integer. If it is ``1", the integer part should be added by ``1", otherwise the integer part is left as it was. The operation is executed by the C-NOT gate and the ``ADDER" module, and its output $|d\rangle$ is the result of the quantum interpolation.

%\addfigure{fig6}{The circuit to realize quantum interpolation}{fig6}{width=0.7\textwidth}

\section{Quantum image rotation}
\label{sec3}

Image rotation is a process of generating another image, through which all the pixels in the image are rotated by a certain angle about a specified point. In this section, the strategy to realize quantum image rotation (viz. QIR) will be proposed based on three-phase shear mappings, i.e. horizontal shear, vertical shear, and a second horizontal shear.

\subsection{Quantum image rotation based on three-phase shear mappings}
\label{sec3-1}

Image rotation can be formulated by the following equation:
\begin{equation}\label{eq6}
\begin{pmatrix}
y_t\\x_t
\end{pmatrix}=
\overbrace{
\begin{pmatrix}
\cos(\theta)&&\sin(\theta)\\-\sin(\theta)&&\cos(\theta)
\end{pmatrix}}^R
\begin{pmatrix}
y_0\\x_0
\end{pmatrix},
\end{equation}
where $y_0$ and $x_0$ represent the pixel position within the original image, while $y_t$ and $x_t$ represent the corresponding position within the rotated image \cite{Gonzalez2010Digital}. $R$ denotes the rotation matrix, $\theta$ represents the rotation angle and its sign indicates the rotation direction, i.e. clockwise (-) and counter-clockwise (+). In order to realize the QIR operation by using the shear mappings, matrix \textit{R} could be rewritten as follows:
\begin{equation}\label{eq7}
\begin{aligned}
R&=
\overbrace{
\begin{pmatrix}
1&&0\\\tan(\theta/2)&&1
\end{pmatrix}^{-1}
\begin{pmatrix}
1&&0\\\tan(\theta/2)&&1
\end{pmatrix}}^{I}
\begin{pmatrix}
\cos(\theta)&&\sin(\theta)\\-\sin(\theta)&&\cos(\theta)
\end{pmatrix}\\
&=
\begin{pmatrix}
1&&0\\\tan(\theta/2)&&1
\end{pmatrix}^{-1}
\begin{pmatrix}
\cos(\theta)&&\sin(\theta)\\-\tan(\theta/2)&&1
\end{pmatrix}\\
&=
\begin{pmatrix}
1&&0\\\tan(\theta/2)&&1
\end{pmatrix}^{-1}
\begin{pmatrix}
\cos(\theta)&&\sin(\theta)\\-\tan(\theta/2)&&1
\end{pmatrix}
\overbrace{
\begin{pmatrix}
1&&0\\\tan(\theta/2)&&1
\end{pmatrix}
\begin{pmatrix}
1&&0\\\tan(\theta/2)&&1
\end{pmatrix}^{-1}}^{I^{-1}} \\
&=
\begin{pmatrix}
1&&0\\\tan(\theta/2)&&1
\end{pmatrix}^{-1}
\begin{pmatrix}
1&&\sin(\theta)\\0&&1
\end{pmatrix}
\begin{pmatrix}
1&&0\\\tan(\theta/2)&&1
\end{pmatrix}^{-1}.
\end{aligned}
\end{equation}

We must carefully address the issue that different pixels from the original image could be assigned the same pixel location \cite{Gonzalez2010Digital}. In other words, we should distribute the multiple output values obtained at the above process to the corresponding output pixel in order to prevent that more than one pixel is located at the same position. Therefore, we use the following matrix (the inverse of matrix \emph{R}):
\begin{equation}\label{eq8}
R^{-1}=
\overbrace{
\begin{pmatrix}
1&&0\\\tan(\theta/2)&&1
\end{pmatrix}}^{M_x}
\overbrace{
\begin{pmatrix}
1&&-\sin(\theta)\\0&&1
\end{pmatrix}}^{M_y}
\overbrace{
\begin{pmatrix}
1&&0\\\tan(\theta/2)&&1
\end{pmatrix}}^{M_x'},
\end{equation}
where $M_x$, $M_y$, and $M_x'$ are  the shear matrices we will use to execute the proposed three shear-based QIR operations. After these steps, the original matrix \textit{R} is transformed to the product of the three sheer matrices, i.e. $M_x$, $M_y$, and $M_x'$, which is presented as:
\begin{equation}\label{eq9}
\begin{pmatrix}
y_t\\x_t
\end{pmatrix}
=
R
\begin{pmatrix}
y_0\\x_0
\end{pmatrix} \Rightarrow
R^{-1}
\begin{pmatrix}
y_t\\x_t
\end{pmatrix}=
\begin{pmatrix}
y_0\\x_0
\end{pmatrix} \Rightarrow
M_xM_yM_x'\begin{pmatrix}
y_t\\x_t
\end{pmatrix}=
\begin{pmatrix}
y_0\\x_0
\end{pmatrix}.
\end{equation}

Hence, we have shown how QIR operation is decomposed into three phases (two-time horizontal shear operations and one vertical shear operation). We now proceed to  discuss the realization of horizontal and vertical shears as unitary circuits. 

\subsection{Shear mapping operations}
\label{sec3-2}

Shear mapping can be either horizontal (i.e. shear parallel to the \emph{x} axis) or vertical (i.e. shear parallel to the \emph{y} axis). Horizontal shear is a function that shifts a original point with coordinates $(x, y)$ to another point located at $[{x+y\times\tan(\theta /2)}, y]$, while vertical shear is a so-called movement from a point $(x, y)$ to another one, i.e. $[x, {y-x\times\sin(\theta)}]$. In the following, the function $\tan(\theta/2)$ is known as the shear factor and $\theta$ as the rotation angle.

We suggest to select the image centroid as the rotation center instead of the default upper left corner. Thus, we are able to divide one quantum image into two halves, i.e. top-bottom halves by horizontal reference line or left-right halves by vertical reference line. Consequently, equations for shear mapping are stated as follows:
\begin{description}
  \item[(\romannumeral1)] Shear top half along the negative direction of \textit{x}-axis ($STH_x^-$):
  \begin{equation}\label{eq10}
  x_s=x-(Y_{mid}-y)\times tan(\theta/2),
  \end{equation}
  \item[(\romannumeral2)] Shear bottom half along the positive direction of \textit{x}-axis ($SBH_x^+$):
  \begin{equation}\label{eq11}
  x_s=x+(y-Y_{mid})\times tan(\theta/2),
  \end{equation}
  \item[(\romannumeral3)] Shear left half along the positive direction of \textit{y}-axis ($SLH_y^+$):
  \begin{equation}\label{eq12}
  y_s=y+(X_{mid}-x)\times sin(\theta),
  \end{equation}
  \item[(\romannumeral4)] Shear right half along the negative direction of \textit{y}-axis ($SRH_y^-$):
  \begin{equation}\label{eq13}
  y_s=y-(x-X_{mid})\times sin(\theta),
  \end{equation}
\end{description}
where $x_s$ and $y_s$ indicate the pixel position after shear mapping, and $X_{mid}$ and $Y_{mid}$ indicate the reference line, i.e. the median of \textit{x}-axis and \textit{y}-axis, respectively. Since pixel points on separated sides around the reference line are displaced towards opposite directions, the image rotation with the image centroid as the rotation center works.

%\addfigure{fig7}{The circuit of horizontal shear to the top half of the quantum image}{fig7}{width=1.00\textwidth}
\begin{figure} [htbp]
%\vspace*{13pt}
\centerline{\includegraphics[width=12cm]{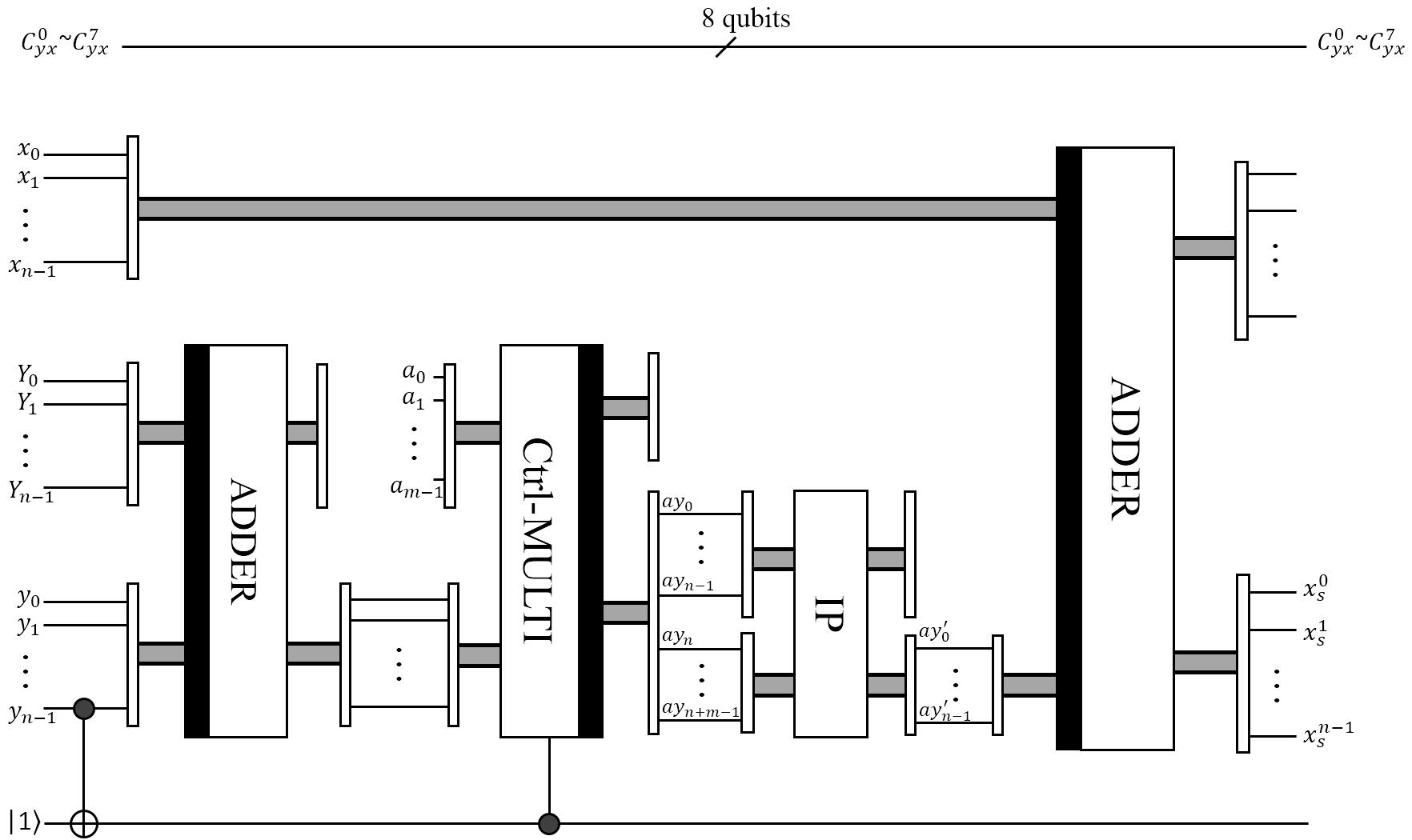}} %100 percent
\vspace*{13pt}
\caption{The circuit of horizontal shear to the top half of the quantum image}
\label{fig7}
\end{figure}

Referring to the $STH_x^-$ in Eq. (\ref{eq10}), Fig. \ref{fig7} completely depicts our proposed solution in terms of a quantum curcuit, in which $|x\rangle$ and $|y\rangle$ represent the coordinate of a pixel before the shear. $|Y\rangle$ represents the horizontal median $Y_{mid}$ of the quantum image and $|a\rangle$ represents the shear factor, i.e. $\tan(\theta/2)$.

Our quantum circuit to compute horizontal shear mapping on the top half quantum image is illustrated as follows:

Step 1: Constrained by a C-NOT gate, the horizontal shear is acted within the top half image when the control qubit is $|0\rangle$. An ADDER module labelled with the left-side black bar is deployed for executing the subtraction operation (cf. Section \ref{sec2-2}). With the subtraction, $|Y_{mid}\rangle-|y\rangle$ is computed correspondingly.

Step 2: With the result in Step 1 as an input, $|a\rangle=|\tan\theta/2\rangle$ is also used as an input to the ``Ctrl-MULTI" module to obtain the multiplication result of $(|Y_{mid}\rangle-|y\rangle)$ and $|\tan(\theta/2)\rangle$. ``Ctrl-MULTI" is short for quantum controlled multiplier which has been introduced in Section \ref{sec2-3}.

Step 3: Usually, the result through Step 2 manifests itself as a binary decimal that cannot reflect the precise position of the displaced pixel. So the ``IP" module, which is short for quantum interpolation operation, is used to round up or down the fraction part and to produce the corresponding integer, i.e. $\lceil(|Y_{mid}\rangle-|y\rangle)|\tan(\theta/2)\rangle\rfloor$ (cf. Section \ref{sec2-4}).

Step 4: Another ``ADDER" module is applied to execute the subtraction operation (where the result obtained in Step 3 is the minuend while the original coordinate value $\vert x\rangle$ is the subtractor) so that the new location of the displaced pixel is achieved.

\begin{figure} [htbp]
%\vspace*{13pt}
\centerline{\includegraphics[width=12cm]{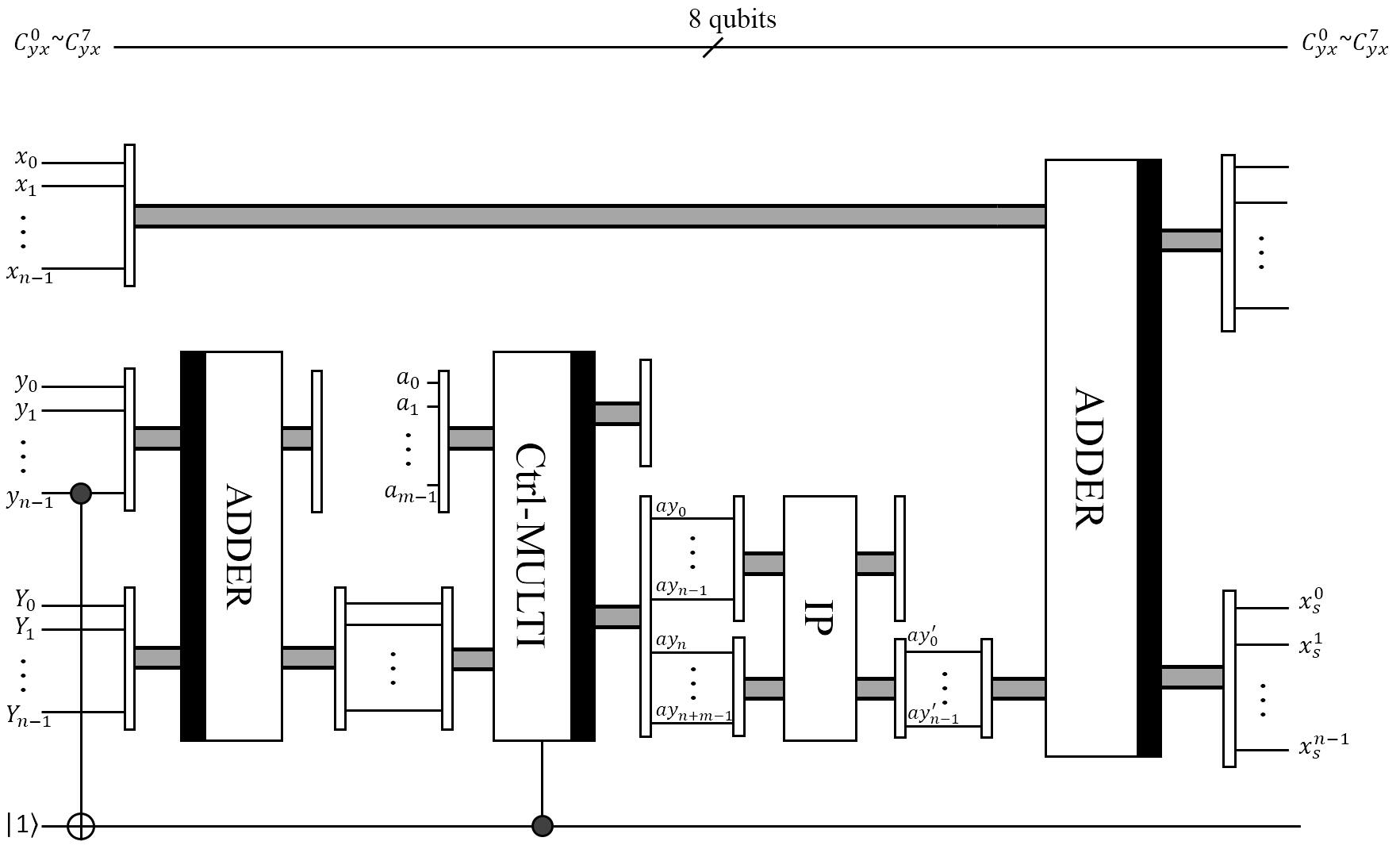}} %100 percent
\vspace*{13pt}
\caption{The circuit to realize horizontal shear to the bottom half of the quantum image}
\label{fig8}
\end{figure}

We now proceed to apply a shear mapping on the pixels at the bottom half of the quantum image. As presented in Eq. (\ref{eq11}), the difference to realize the $SBH_x^+$ is to calculate the $\vert y\rangle-\vert Y_{mid}\rangle$ firstly by using the subtraction module in the circuit when the control qubit is $|1\rangle$. In the final operation, the quantum adder module is used to compute the addition of $\vert x\rangle$ and $(|y\rangle-|Y_{mid}\rangle)|tan(\theta/2)\rangle$. The circuit implementing the procedure is shown in Fig. \ref{fig8}. After these two phases, the horizontal shear to a quantum image is achieved by taking the image centroid as the shear center.

%\addfigure{fig8}{The circuit to realize horizontal shear to the bottom half of the quantum image}{fig8}{width=1.00\textwidth}

\begin{figure} [htbp]
%\vspace*{13pt}
\centerline{\includegraphics[width=12cm]{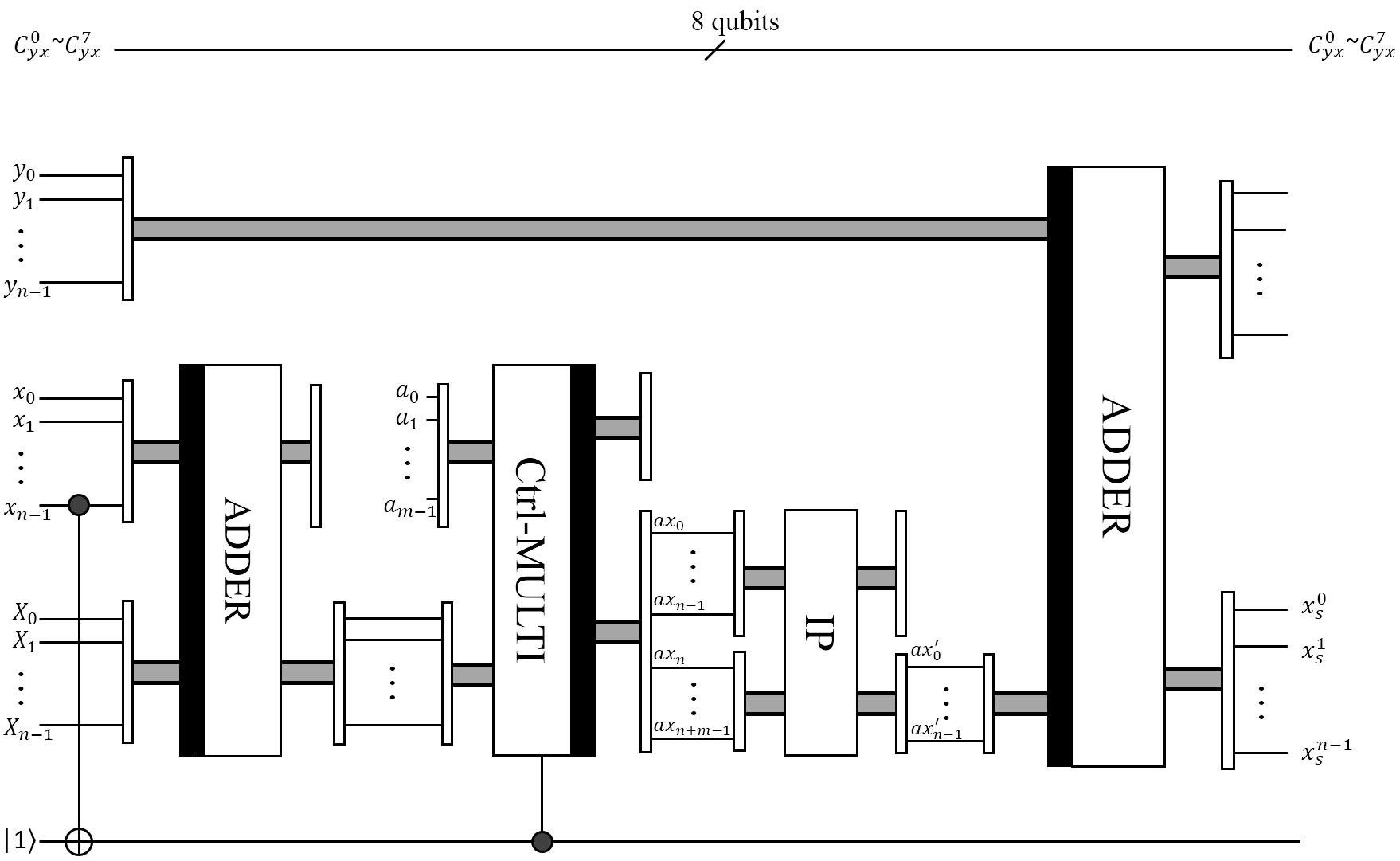}} %100 percent
\vspace*{13pt}
\caption{The circuit to realize vertical shear to the left half of the quantum image}
\label{figx}
\end{figure}

Similar to horizontal shears, to make a vertical shear centered at the image centroid requires to divide the quantum image into two halves, i.e. left half and right half.  The circuit that implements $SLH_y^+$ (Eq. (\ref{eq12}))  is presented on Fig. \ref{figx}, where $|x\rangle$ and $|y\rangle$ represent the coordinate of a pixel before the shear, $|X\rangle$ represents the vertical median $X_{mid}$ of the quantum image and $|a\rangle$ represents the shear factor (in this case,  $\sin(\theta)$). The qubit of $x_{n-1}$ is used to determine  which vertical half of the quantum image (left half for $x_{n-1}=0$ or right half for $x_{n-1}=1$) the vertical shear transformation will be applied on. The ``ADDER" module with left-side bar is used to execute the operation ``$X_{mid} - x$" and the result is used as one of the two inputs of multiplier, while the other input of multiplier is $\vert sin(\theta)\rangle$ which is labelled as $\vert a\rangle$ in the circuit. By utilizing the ``IP" module to convert the decimal number from the multiplier to an integer, the distance of the displacement is added to the original coordinate value $\vert y\rangle$ through the ``ADDER" module so that the new location of the displaced pixel is computed.

%\addfigure{figx}{The circuit to realize vertical shear to the left half of the quantum image}{figx}{width=1.00\textwidth}

\begin{figure} [htbp]
%\vspace*{13pt}
\centerline{\includegraphics[width=12cm]{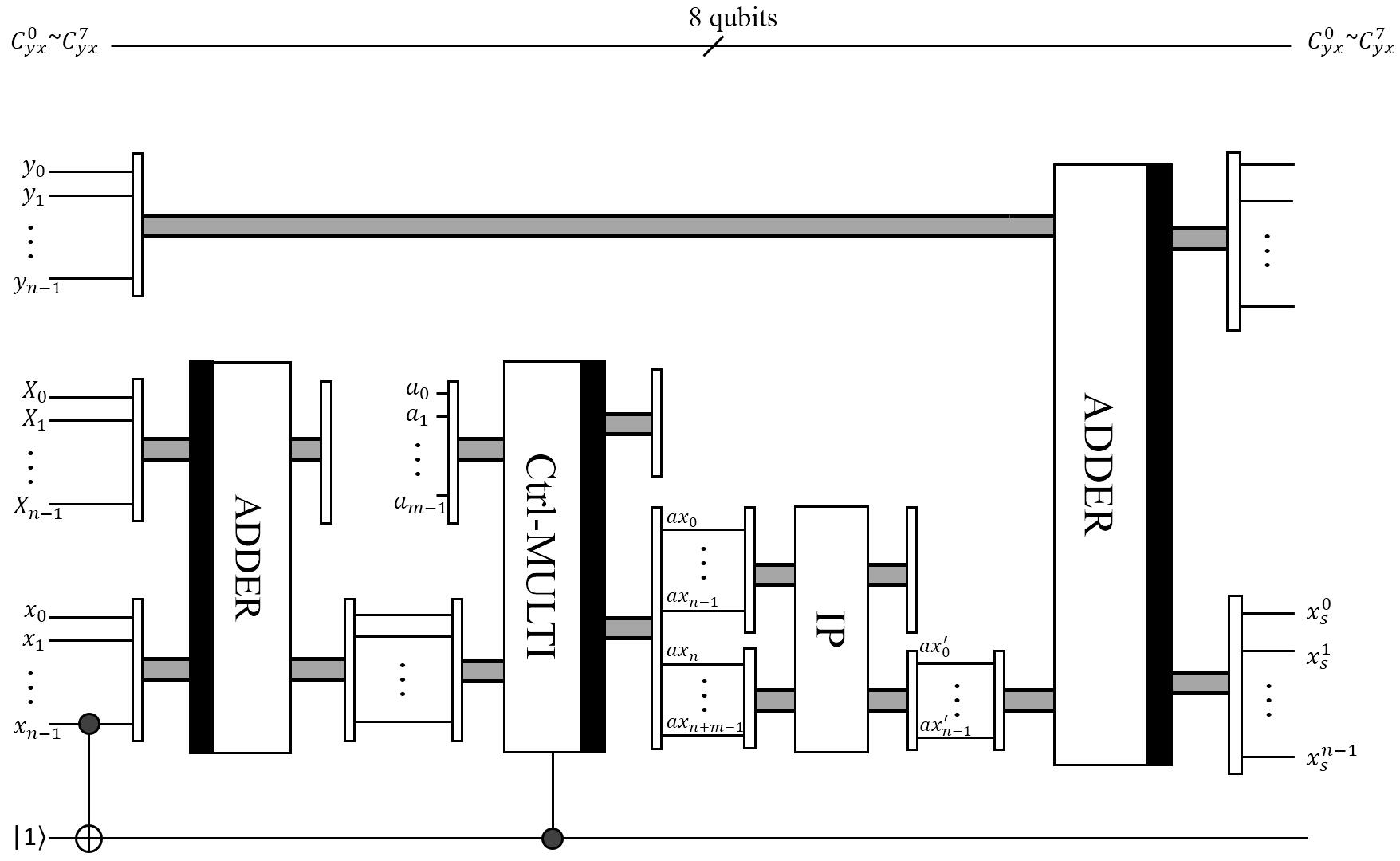}} %100 percent
\vspace*{13pt}
\caption{The circuit to realize vertical shear to the right half of the quantum image}
\label{figy}
\end{figure}

In order to complete the definition of vertical shear mapping, we need to further perform the shear mapping to the right part of quantum image according to Eq.(\ref{eq13}). Fig. \ref{figy} presents the circuit of vertical shear to the right part of quantum image. The main differences between the shear of left half and right half are two-fold: First, as discussed earlier, the ``ADDER'' with left-side bar is a subtraction operation, yet in this turn, $\vert x\rangle$ is treated as the minuend and $\vert X_{min}\rangle$ becomes the subtrahend; Second, another subtraction operation is used at the end of the circuit to obtain the final location of the displaced pixel, i.e. $\vert y\rangle-\lceil (\vert x\rangle-\vert X_{mid}\rangle)\vert \sin(\theta)\rangle \rfloor$.

%\addfigure{figy}{The circuit to realize vertical shear to the right half of the quantum image}{figy}{width=1.00\textwidth}

\begin{figure} [htbp]
%\vspace*{13pt}
\centerline{\includegraphics[width=12cm]{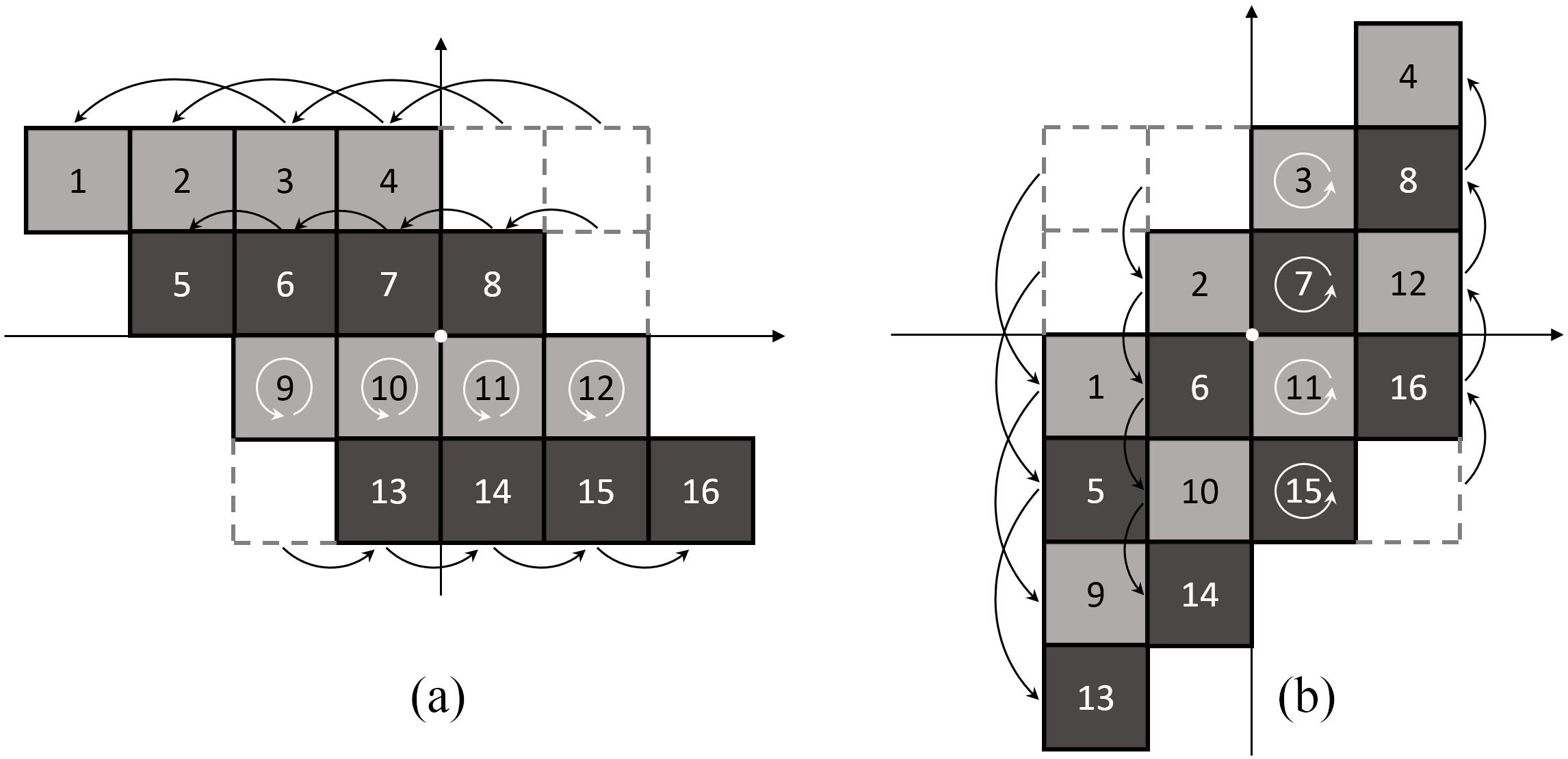}} %100 percent
\vspace*{13pt}
\caption{\textbf{A $4\times 4$ NEQR image to illustrate the horizontal and vertical shear}}
\label{figz}
\end{figure}

To illustrate the procedure of horizontal and vertical shear, a $4\times 4$ NEQR image with different colors for each row is  presented in Fig. \ref{figz}. In Fig. \ref{figz}(a), the image is divided into two halves by the horizontal axis, in which the top two rows are sheared to the negative direction while the bottom two rows are sheared to the positive direction (in this case, the shear factor is $\tan45^\circ=1$). As presented in Eqs.(\ref{eq10},\ref{eq11}), the displacements of each row are 2, 1, 0, and 1 in that order. The arrowhead and tail indicate the direction and distance of the displacement. The ``dashed boxes'' indicate the vacated locations after the pixels moved out. Turn discussion to Fig. \ref{figz}(b), the image is divided into two halves by the vertical axis, in which the left two columns are sheared to the negative direction while the right two columns are sheared to the positive direction. The depiction and illustration of Fig. \ref{figz}(b) is similar to those of Fig. \ref{figz}(a).

%\addfigure{figz}{\textbf{A $4\times 4$ NEQR image to illustrate the horizontal and vertical shear}}{figz}{width=0.85\textwidth}

\subsection{Analysis of the network complexity}
\label{sec3-3}

The complexity of quantum image processing algorithms is usually estimated in terms of the number of elementary quantum gates \cite{Yan2016A} and, in this paper, we take the C-NOT gate as the basic unit for estimating the complexity of our quantum circuits.
If the size of quantum image is $2^n\times2^n$, the input scale of the network is $O(n)$. The complexities of the operations used in the quantum shear mapping are as follows:

\begin{description}
  \item[(\romannumeral1)] Quantum self-adder: $n$;
  \item[(\romannumeral2)] Quantum adder: $28n-12$;
  \item[(\romannumeral3)] Quantum interpolation: $28n-11$;
  \item[(\romannumeral4)] Quantum control multiplier: ${14.5}n(n+2m-1)$.
\end{description}

It is noteworthy that the Toffoli gate can be expressed as a quantum circuit composed of  $6$ C-NOT gates \cite{Vedral1996Quantum}, hence the complexity of Toffoli gate is quantified as $6$. The complexities of the quantum adder has been discussed in Section \ref{sec2-2}. In addition, the quantum interpolation includes one C-NOT gate and one adder as shown in Fig. \ref{fig6}. Since the quantum control multiplier is constructed by $n$-step adder modules, each adder module is configured with a self-adder module, and the input scale of both the adder and the self-adder modules can be specified as $m+i$ at the $i^{\text{th}}$ step, the concerned complexity could be described correspondingly.

Thus, it is possible to estimate the complexity of the network used for computing a horizontal shear mapping onto the top half quantum image, which is:  
\begin{equation}\label{eq14}
\begin{aligned}
&\overbrace{
(28n-12)\times2}^{two\ adders}
+
\overbrace{
29mn+29(n^{2}-n)/2}^{control\ multiplier}
+
\overbrace{
28n-11}^{interpolation}\\
&
=
({14.5}n+29m+{69.5})n-35
.
\end{aligned}
\end{equation}

Meanwhile, the network for horizontal shear onto the bottom half quantum image could be parsed into two adders (wherein one is for subtraction operation), one control multiplier, and one interpolation. So its complexity is equivalent to its counterpart on the top half.

Through these two procedures, the horizontal shear onto the quantum image is performed and its corresponding complexity is given by $29n^2+58mn+139n-70$. As elaborated above, the vertical shear resembles its horizontal counterpart other than switching the \emph{x}-axis and \emph{y}-axis, hence the complexity of vertical shear is the same as that of the horizontal shear. To complete the quantum image rotation, three-phase shear mappings should be operated along the whole computing process.

\section{Experiments on quantum image rotation}
\label{sec4}

In the experiment, we select to use one $512\times512$ NEQR encoding Lena image as the original sample image. According to Eq.(\ref{eq1}), 7 qubits are employed to encode the position information and 8 qubits to encode the color information. In total, 15 qubits are configured to represent the above-mentioned NEQR sample image.

\begin{figure} [htbp]
%\vspace*{13pt}
\centerline{\includegraphics[width=12cm]{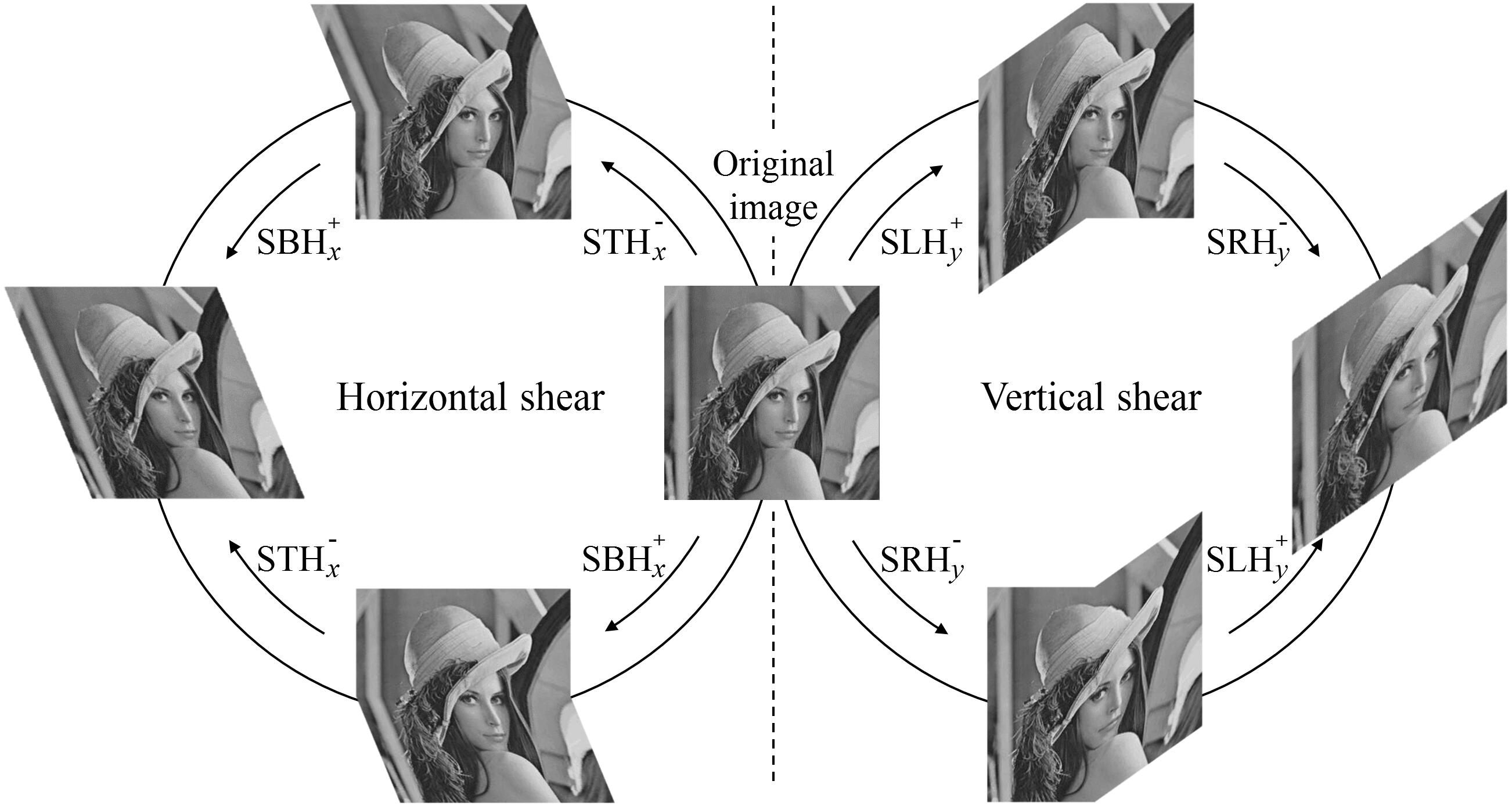}} %100 percent
\vspace*{13pt}
\caption{Strategies of horizontal shear and vertical shear}
\label{fig9}
\end{figure}

Along three-phase shear mappings for the QIR operation, both of the half images are sheared towards the opposite directions so as to make the quantum image rotated around the image centroid. With the original quantum Lena image, there are two potential ways of shear mappings, i.e. horizontal shear and vertical shear. To operate the horizontal shear, we could use the shear top half (STH) first and then perform the shear bottom half (SBH), and vice versa. In the similar way, to operate the vertical shear, we could execute shear left half (SLH) first and then invoke shear right half (SRH) afterwards, and vice versa, as presented in Fig.~\ref{fig9}.

%\addfigure{fig9}{Strategies of horizontal shear and vertical shear}{fig9}{width=0.90\textwidth}

%verify the feasibility and effectiveness of the contributed work, 

\begin{figure} [htbp]
%\vspace*{13pt}
\centerline{\includegraphics[width=12cm]{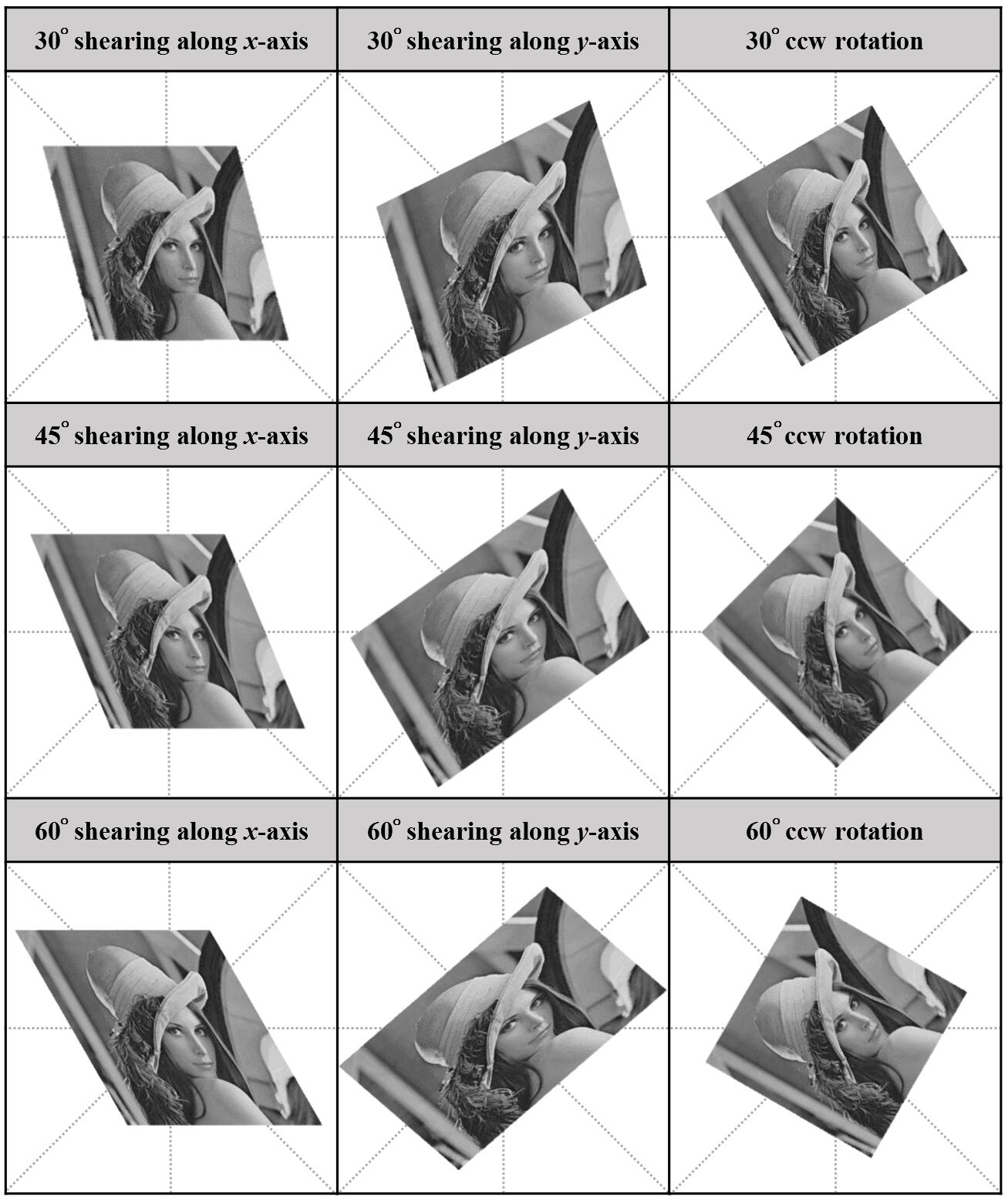}} %100 percent
\vspace*{13pt}
\caption{The quantum image rotation based on three-phase shear mappings with different angles}
\label{fig10}
\end{figure}

We now present results of $30^{\circ}$, $45^{\circ}$ and $60^{\circ}$ counter-clockwise image rotations, respectively. In each case, three-phase shear procedures will be run and results of shear mappings and rotation are presented in Fig. \ref{fig10}. Each row of Fig. \ref{fig10} illustrates the three-phase shear procedures (i.e., horizontal shear, vertical shear, and a second horizontal shear) that rotate the quantum image by the corresponding angles (rotated images are shown on the third column).

%\addfigure{fig10}{The quantum image rotation based on three-phase shear mappings with different angles}{fig10}{width=0.90\textwidth}

As indicated in the experimental results in Fig. \ref{fig10}, utilizing the horizontal shear and the vertical shear separately to rotate the quantum image ensures that each interpolation only concerns two adjacent pixel points along either \emph{x}-axis or \emph{y}-axis, which enables to gain the lower computational complexity compared with the simultaneous interpolations concerning two-dimensional plane. In addition, the shear factor in each row or column is invariant (i.e., $\sin 30^\circ$, $\sin 45^\circ$, and $\sin 60^\circ$ in the vertical shear) so that the relative location of the pixels in each row or column are preserved. Therefore, the problems of blocking and/or blurring could be avoided during the rotation.

\section{Conclusions}
\label{sec5}
In this paper, we have proposed a method of quantum image rotation (QIR) that consists of employing shear mapping operations successively on NEQR quantum images. To compute horizontal and vertical shear mappings for implementing an arbitrary rotation, we employ the following circuits: quantum self-adder, quantum controlled multiplier, and quantum interpolation. The contributions in this study mainly include the proposal of QIR operations by using the three-phase horizontal and vertical shear, the partition strategy that selects to divide image at the median line so as to make the rotation occur around the image centroid, and the design of quantum controlled multiplier with quantum self-adder to process any pair of qubit numbers (the coordinate and the shear factor in the case) iteratively. In addition, we also presented a quantum interpolation circuit as a convenient approach of covering the fractional part (i.e., round-off) to determine the new position of the displaced points.

\begin{acknowledgements}
This work is supported by the National Natural Science Foundation of China (No. 61502053) and the Science \& Technology Development Program of Jilin Province, China (No. 20170520065JH). SEVA gratefully acknowledges the financial support of Tecnol\'{o}gico de Monterrey, Escuela de Ciencias e Ingenier\'{\i}a and CONACyT (SNI member number 41594 as well as Fronteras de la Ciencia project No. 1007).
\end{acknowledgements}

\end{document}